\documentclass[
  journal=pasa,
  manuscript=article-type,
  year=2020,
  volume=37,
]{cup-journal}

\usepackage{amsmath}
\usepackage[nopatch]{microtype}
\usepackage{booktabs}

\title{Effects of angular scattering and H+p, H+H collisions on the properties of interstellar atoms in the heliosphere}

\author{A.V. Titova}
\affiliation{Space Research Institute, Russian Academy of Sciences, Profsoyuznaya Str. 84/32, Moscow, 117335 Russia}
\alsoaffiliation{Lomonosov Moscow State University, Moscow Center for Fundamental and Applied Mathematics, GSP-1, Leninskie Gory, Moscow, 119991 Russia}
\alsoaffiliation{Faculty of Physics, HSE University, 20 Myasnitskaya Ulitsa, Moscow 101000, Russia}
\email[A. Titova]{avtitova@hse.ru}

\author{V.V. Izmodenov}
\affiliation{Space Research Institute, Russian Academy of Sciences, Profsoyuznaya Str. 84/32, Moscow, 117335 Russia}
\alsoaffiliation{Lomonosov Moscow State University, Moscow Center for Fundamental and Applied Mathematics, GSP-1, Leninskie Gory, Moscow, 119991 Russia}
\alsoaffiliation{Faculty of Physics, HSE University, 20 Myasnitskaya Ulitsa, Moscow 101000, Russia}

\author{S.D. Korolkov}
\affiliation{Space Research Institute, Russian Academy of Sciences, Profsoyuznaya Str. 84/32, Moscow, 117335 Russia}
\alsoaffiliation{Lomonosov Moscow State University, Moscow Center for Fundamental and Applied Mathematics, GSP-1, Leninskie Gory, Moscow, 119991 Russia}

\keywords{heliosphere, interstellar atoms, charge transfer, interstellar medium, stellar-interstellar interactions} 

\begin{document}

\begin{abstract}
Interstellar hydrogen atoms (H atoms) penetrate into the heliosphere through the region of the solar wind interaction with the interstellar plasma due to their large mean free path. Resonant charge exchange of H atoms with protons has been considered as the main interaction process between the components. In the majority of models, other processes like elastic H-H and H-p collisions are not included.  Moreover, it has been assumed that the velocities of the colliding particles remain unchanged during charge exchange. This corresponds to the scattering on the angle of $\pi$ in the centre mass rest frame.

The goal of  this paper is to explore effects of the elastic H-H and H-p collisions as well as the  angular scattering during charge exchange on the distribution of the interstellar atoms in the heliosphere and at its boundary. 

We present results of simple (and therefore, easily repeatable) kinetic model of the interstellar atom penetration through the region of the solar and interstellar winds interaction into the heliosphere. As a result of the model we compute the distribution function of the interstellar atoms at different heliospheric distances. Further, this distribution function is used to compute its moments and potentially observable features such as absorption and backscattered spectra in the Lyman-alpha line. 

Results show that there are differences in the behavior of the distribution function when considering elastic collisions and the changes in the moments of the distribution achieve 10\%. Therefore, in cases where precise calculation of H atom parameters is essential, such as in the modeling of backscattered Lyman-$\alpha$ emission, elastic collisions must be considered.
\end{abstract}

\section{Introduction} \label{sec: introduction}

The solar wind collides with the charged components of the local interstellar medium (LISM) forming so-called heliospheric boundary layer or heliosheath that is bordered by the two shocks - the heliospheric termination shock and the outer bow shock. The formation of the bow shock, however, can be prevented or reduced by the influence of the local interstellar magnetic field. The heliopause is the tangential discontinuity located between the shocks that separates the solar wind and interstellar flows. The first two-shock model of the solar wind/interstellar medium interaction was proposed by \cite{baranov_1970}. Already in 1971 \citeauthor{Bertaux1971} and \citeauthor{Thomas1971} have shown by analysing OGO-5 backscattered Lyman-$\alpha$ observations that interstellar neutral hydrogen penetrates deep into the heliosphere, as close to the Sun as 2-3 AU. \cite{wallis1975} argued that charge exchange with protons in the outer heliosheath between the heliopause and the bow shock prevents interstellar atoms to penetrate into the heliosphere, so only a fraction of atoms may penetrate through.

While passing through the interstellar medium and heliospheric boundaries, light from stars is absorbed in the Lyman-$\alpha$ line. This absorption depends on the column density of the H atoms. In case of nearby stars, when the interstellar absorption is not that broad, the absorption from the heliospheric boundaries -- particularly the hydrogen wall, where H atoms are heated and decelerated -- can be distinguished. This absorption has been recognized as a valuable remote diagnostic tool for studying the heliospheric interface.

The hydrogen wall was first identified by \cite{Linsky_1996}, who analyzed high-resolution ultraviolet absorption spectra of $\alpha$ Cen A and $\alpha$ Cen B obtained with the Goddard High Resolution Spectrograph (GHRS) on the Hubble Space Telescope (HST). Later studies expanded these observations to include more nearby stars, such as Sirius with GHRS (\cite{izmodenov_1999}), as well as 36 Oph (\cite{Wood_2000}), 70 Oph, $\xi$ Boo, 61 Vir, and HD 165185 (\cite{Wood_2005}) using the Space Telescope Imaging Spectrograph (STIS), and other stars (\cite{Wood_2007}).

As it was discussed above, the interstellar H atoms inside the heliosphere were found by studying the backscattered Lyman-$\alpha$ emission. In the early 1970s, the OGO-5 spacecraft provided measurements of backscattered Lyman-$\alpha $ emission outside the geocorona (\cite{Bertaux1971, Thomas1971}). These observations helped confirm the source of the emission. Due to the parallax effect, it was  established that the measured Lyman-$\alpha$ emission originated from interstellar H) atoms flowing into the Solar system from the LISM.

From the late 1970s to the 1980s, significant advancements came from Soviet missions such as Mars-7 (\cite{Bertaux_1976}) and Prognoz-5/6 (\cite{Bertaux_1977, Bertaux_1985, Lallement_1984, Lallement_1985}). These spacecraft carried photometers equipped with hydrogen absorption cells, enabling studies of both the intensity and spectral properties of the radiation. Backscattered Lyman-$\alpha$ emission was also measured in the outer heliosphere by Voyager Ultraviolet Spectrometer (UVS). Since the 1990s, the Lyman-$\alpha$ intensity has been measured on the Solar and Heliospheric Observatory (SOHO)/Solar Wind ANisotropy (SWAN) providing a complete map of the sky in about 24 hours (\cite{Bertaux_1995}). The backscattered Lyman-$\alpha$ emission spectra
 were observed by the GHRS and STIS on board the HST (\cite{Clarke_1995, Clarke_1998}).

To model the backscattered Lyman-$\alpha$ emission, it is essential to know the distribution function of the H atoms everywhere inside the heliosphere, as the intensity depends on the projection of the distribution function along the line of sight. Once this distribution is determined, a radiative transfer equation must be solved. The modeling of the H atom velocity distribution function inside the heliosphere and the resulting backscattered emission is discussed in, for example, \cite{KATUSHKINA2011, Kubiak_2021}.

The mean free path of interstellar atoms in the heliosheath is comparable to the size of the region. It makes self-consistent (including charge exchange) modeling of the two-component (plasma and H atoms) LISM interaction with the solar wind quite complex. Rigorous approach requires to solve gas-dynamic (or MHD) equations for charged components together with kinetic equation for neutral components. The first self-consistent kinetic-gasdynamic model was developed by \cite{baranov1993}. It was shown that plasma and neutral component significantly change their momentum and energy due to charge exchange. The transfer of momentum leads to essential displacements of the termination shock, heliopause and bow shock locations toward the Sun (see  \cite{izmodenov2000}). A recent study by \cite{korolkov2024} investigated the effects of charge exchange with interstellar hydrogen atoms on plasma flow in heliospheric and astrospheric shock layers.

In addition to charge exchange, at collision energies in the center-of-mass frame less than 10 keV the dominant (in terms of the total cross section) process between atoms and protons is elastic scattering. Moreover, hydrogen atoms can also interact with each other through H-H elastic collisions. The effects of charge exchange on the global structure of the heliosphere and atoms distribution have been well explored, however, the question of the necessity to take into account these elastic collisions in theoretical models of the global heliosphere remains open.

\cite{williams1997} have suggested that a population of hot hydrogen atoms is created in the heliosphere through elastic H-H collisions between energetic solar atoms (neutralized solar wind) and interstellar atoms. However, \cite{izmodenov2000_HP} argued that this effect will be negligibly small because the momentum cross section for elastic H-H collisions is several order of magnitudes smaller than the charge exchange cross section. Calculated momentum cross section of elastic collisions has been presented in this paper.

\cite{heerikhuisen2009} checked the consequences of angular scattering in charge exchange collisions for the global modeling of the heliosphere. A few runs using several types of collisional cross sections were performed. It was shown that even with the isotropic scattering (which assumes a uniform differential cross-section) only changes the global solution slightly. Thus, it was concluded that angular scattering does not significantly influence the global distribution of plasma.

\cite{swaczyna2019} applied the differential H-p cross section by \cite{schultz_2016} to determine the impact of the momentum transfer due to angular scattering in charge exchange collisions on distributions of the secondary populations of interstellar hydrogen atoms. The initial hydrogen atoms distribution was considered maxwellian with different temperatures ranging from 7500 to 22,500 K and relative velocities less than 50 km/s.  Authors claimed that the momentum transfer leads to the increase of secondary population velocities in the direction of motion of the primary population and the heating of secondary population by up to $\sim$3000 K.

Subsequent research continued by taking the influence of H-p elastic collisions into account. \cite{rahmanifard2023} simulated the transport of hydrogen atoms through the outer heliosheath with angular charge exchange and elastic collisions between atoms and protons. It was shown that angular scattering by both charge exchange and elastic collisions decelerates and heats the primary population. However, the secondary population does not significantly change.

The current state of knowledge on elastic H-H and H-p collisions appears contradictory. Different approaches and approximations lead to different outcomes. Therefore, in order to understand whether it is needed to include H-p and H-H elastic collisions in global numerical models of the heliosphere we decided to conduct our own research. In this paper, we present a simple kinetic model of the interstellar atoms penetration through the region of the solar and interstellar winds
interaction into the heliosphere. As a result, the distribution function of hydrogen atoms and its moments were calculated at various heliocentric distances. Additionally, based on the obtained distributions the observable properties such as Lyman-$\alpha$ backscattered solar emission and
absorption spectra were found.

The rest of the paper is organised as follows: Section \ref{sec: model} describes the model used in the paper, Section \ref{sec: results} presents the results of the calculations and is divided into three parts dedicated to distribution function, moments and observable spectra. We conclude in Section \ref{section: conclusions}. The detailed information about the cross sections used in the paper is presented in \ref{appendix: cross section}.

\begin{figure}[t]
    \center{\includegraphics[width=1\linewidth]{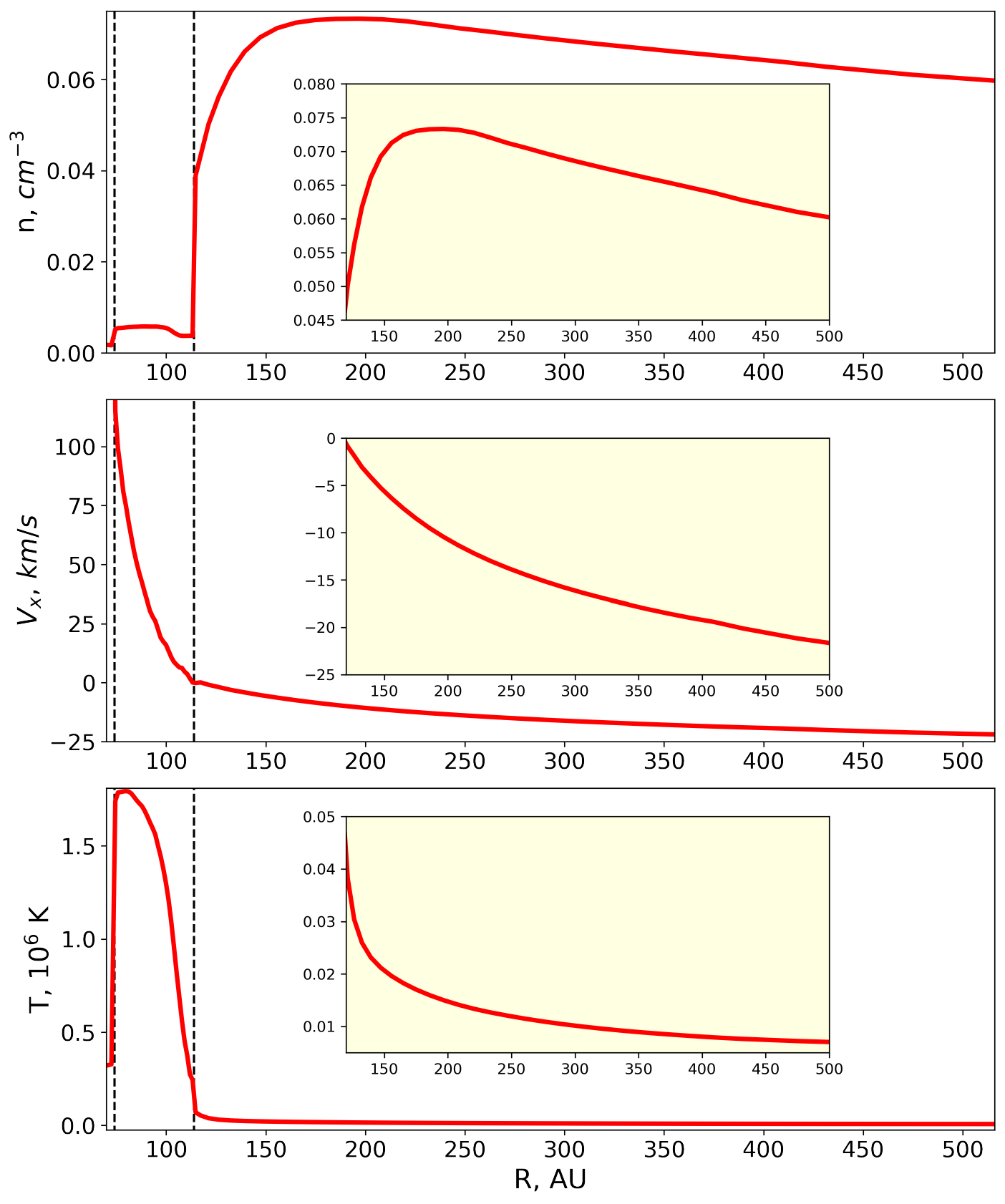}}
    \caption{The distributions of plasma parameters along the upwind direction from the heliospheric model by \cite{izmodenov_2020}. The vertical dotted black lines match the discontinuity surfaces of the heliospheric model: the heliospheric termination shock (TS, 75 au), the heliopause (HP, 115 au). Boundary conditions: $n_{H,LISM}=0.14$ cm$^{-3}$,  $n_{p, LISM} = 0.04$ cm$^{-3}$, $n_{He^{+}, LISM} = 0.003$ cm$^{-3}$, $V_{LISM} = 26.4$ km/s, $T_{LISM} = 6530$ K, $B_{LISM} = 3.75$ $\mu$G, $\alpha=60^{\circ}$, $n_{p, E} = 5.94$ cm$^{-3}$, $V_{R,E} = 432.4$ km/s}
    \label{figure: plasma_parameters}
\end{figure}

\section{Model} \label{sec: model}

The model considered in this section is stationary, one-dimensional in the physical space, and three-dimensional in the velocity space. The space coordinate $x$ can be considered the distance in the upwind direction, i.e. the positive direction of the x-axis is directed towards the undisturbed interstellar flow. In order to obtain the velocity distribution function $f_H(\boldsymbol{v}, x)$ we are solving the following Boltzmann kinetic equation:
\[
v_{x} \cdot \frac{\partial f_H}{\partial x}=I_{ex}+I_{p}+I_{H},\]
where
\begin{eqnarray*}
I_{ex}=\int_0^{\pi} \int_0^{2 \pi}\int_{-\infty}^{+\infty} \int_{-\infty}^{+\infty} \int_{-\infty}^{+\infty} (f_p(\boldsymbol{w}', x)  f_H(\boldsymbol{v}', x) -\nonumber \\
-f_p(\boldsymbol{w}, x) f_H(\boldsymbol{v}, x) )   g \frac{d\sigma_{ex}(g, \chi)}{d  \Omega}sin \chi d \chi d \varepsilon  d \boldsymbol{w}, 
\end{eqnarray*}
\begin{eqnarray*}
I_{p}= \int_0^{\pi} \int_0^{2 \pi} \int_{-\infty}^{+\infty} \int_{-\infty}^{+\infty} \int_{-\infty}^{+\infty} (f_p(\boldsymbol{w}', x)  f_H(\boldsymbol{v}', x) -  \nonumber \\
    -f_p(\boldsymbol{w}, x) f_H(\boldsymbol{v}, x) )  
    g \frac{d\sigma_{p}(g, \chi)}{d  \Omega} sin \chi d \chi d \varepsilon d \boldsymbol{w},
\end{eqnarray*}
\begin{eqnarray*}
I_{H}= \int_0^{\pi} \int_0^{2 \pi} \int_{-\infty}^{+\infty} \int_{-\infty}^{+\infty} \int_{-\infty}^{+\infty} (f_H(\boldsymbol{w}', x)  f_H(\boldsymbol{v}', x) -  \nonumber \\
    -f_H(\boldsymbol{w}, x) f_H(\boldsymbol{v}, x) )
    g \frac{d\sigma_{H}(g, \chi)}{d  \Omega} sin \chi d \chi d \varepsilon d \boldsymbol{w}.
\end{eqnarray*}

Here $f_H= f_H(\boldsymbol{v}, x)$ and  $f_p= f_p(\boldsymbol{w}, x)$ are the velocity distribution functions of atomic hydrogen and protons respectively, and $\frac{d\sigma_{ex}(g, \chi)}{d  \Omega}, \frac{d\sigma_{p}(g, \chi)}{d  \Omega}, \frac{d\sigma_{H}(g, \chi)}{d  \Omega}$ are the differential cross sections of charge-exchange, elastic collisions with protons, and elastic H-H collisions;
 $g = |\boldsymbol{v} - \boldsymbol{w} |$ is the relative velocity of the colliding particles, 
$\chi$ is the scattering angle in the center-of-mass frame,  $\varepsilon$ is the angle that defines the plane of the interacting particles; $\boldsymbol{v}' = \boldsymbol{v}'(\boldsymbol{v}, \boldsymbol{w}, \chi)$,  $\boldsymbol{w}'=\boldsymbol{w}'(\boldsymbol{v}, \boldsymbol{w}, \chi)$ are the velocities of atom and protons after the collision. These velocities are defined from the momentum and energy conservation laws in the interaction as follows. 

In the center-of-mass frame the absolute value of the relative velocity does not change after the collision. The relative velocity in the center-of-mass frame is $\boldsymbol{{g}_{CM}} = \boldsymbol{v}- \frac{\boldsymbol{v} + \boldsymbol{w} }{2}$ (protons and atoms are considered to have equal mass).
The relative velocity after the collision becomes
$
\boldsymbol{{g}_{CM}'}=|g_{CM}|\cdot (cos(\chi)\boldsymbol{e}_1+sin(\chi)cos(\varepsilon)\boldsymbol{e}_2+sin(\chi)sin(\varepsilon)\boldsymbol{e}_3),
$
where $\boldsymbol{e}_1$ is a unit vector collinear to $\boldsymbol{g}$, $\boldsymbol{e}_2$ and $\boldsymbol{e}_3$ are vectors perpendicular to $\boldsymbol{e}_1$.
In the laboratory frame the atom velocity after collision can be found as
$
    \boldsymbol{v}' = \boldsymbol{{g}_{CM}'} +  \frac{\boldsymbol{v} + \boldsymbol{w}}{2}.
$

The distribution function $f_p(w_{x}, w_{y}, w_{z}, x)$ is assumed to be known and maxwellian
\[
f_p (w_{x}, w_{y}, w_{z}, x) =  \frac{n_p(x)}{c_p^3\pi \sqrt{\pi}} exp \left(-\frac{(w_{x} - U_p(x))^2+w_{y}^2+w_{z}^2}{c_p^2} \right), 
\]
Here
$
c_p = \sqrt{\frac{2 k_B T_p(x)}{ m_p}} 
$
is the thermal velocity, $k_B$ is the Boltzmann constant, $m_p$ is the proton mass.
$n_p(x), U_p(x), T_p(x)$ are the number density, bulk velocity, and temperature of protons along the upwind direction. In our calculations, we used the distributions of these parameters obtained in the frame of \cite{izmodenov_2020} model (see, Figure \ref{figure: plasma_parameters}).

To finish the formulation of the problem, we set the boundary conditions for $f_H$ at $x_0 \approx 500$ AU, where the interstellar medium is assumed to be undisturbed: 

\begin{eqnarray*}
f_H(v_{x}, v_{y}, v_{z}, x_0)=  \frac{n_{H, \infty}}{ c_{H, \infty}^{3}\pi \sqrt{\pi}} exp \left(-\frac{(v_{x} - U_{H, \infty})^2+v_{y}^2+v_{z}^2}{c^2_{H,\infty}} \right),\nonumber \\
 v_{x} <0.
\end{eqnarray*}

Here $
c_{H,\infty} = \sqrt{\frac{2 k_B T_{H,\infty}}{ m_H}}$ is the thermal velocity.
In our calculations we assume $n_{H, \infty} =$0.14 cm$^{-3}$, $U_{H, \infty}$ =26.4 km/s, $T_{H,\infty}$ = 6530 K. Exact values are not important for the presented study.

For H-p charge exchange and elastic collisions we solve the problem using the imitative Monte Carlo scheme which is described in detail in \cite{malama1991}. 

For H-H collisions the Boltzmann collision integral is non-linear. To calculate it properly the iterative method is implied. In the first step only interaction with protons is taken into account. The 3D distribution function $f_H$ is calculated using the Monte Carlo method and stored at every point in the computational mesh. We compute the distribution functions numerically, without assuming any specific shape or making any approximations. In step 2, we take the H-H collisions into account and run trajectories of H particles as it is done as usual in the test particle Monte Carlo code assuming
that the local H atom distribution is known from the previous step. In subsequent steps, we update the distribution function of particle-partners in collisions and repeat the run of test particles.
The iteration procedure is finished when the distribution functions of the two consecutive iterations become indistinguishable.

For the calculation of the one-dimensional distribution function the $v_x$ velocity space is divided into spatial cells of equal size of $\sim$0.8 km/s. The number of launched particles is $N \sim 10^{6}-10^{7}$.
\begin{figure}[t]
    \centering
    \includegraphics[width=1\linewidth]{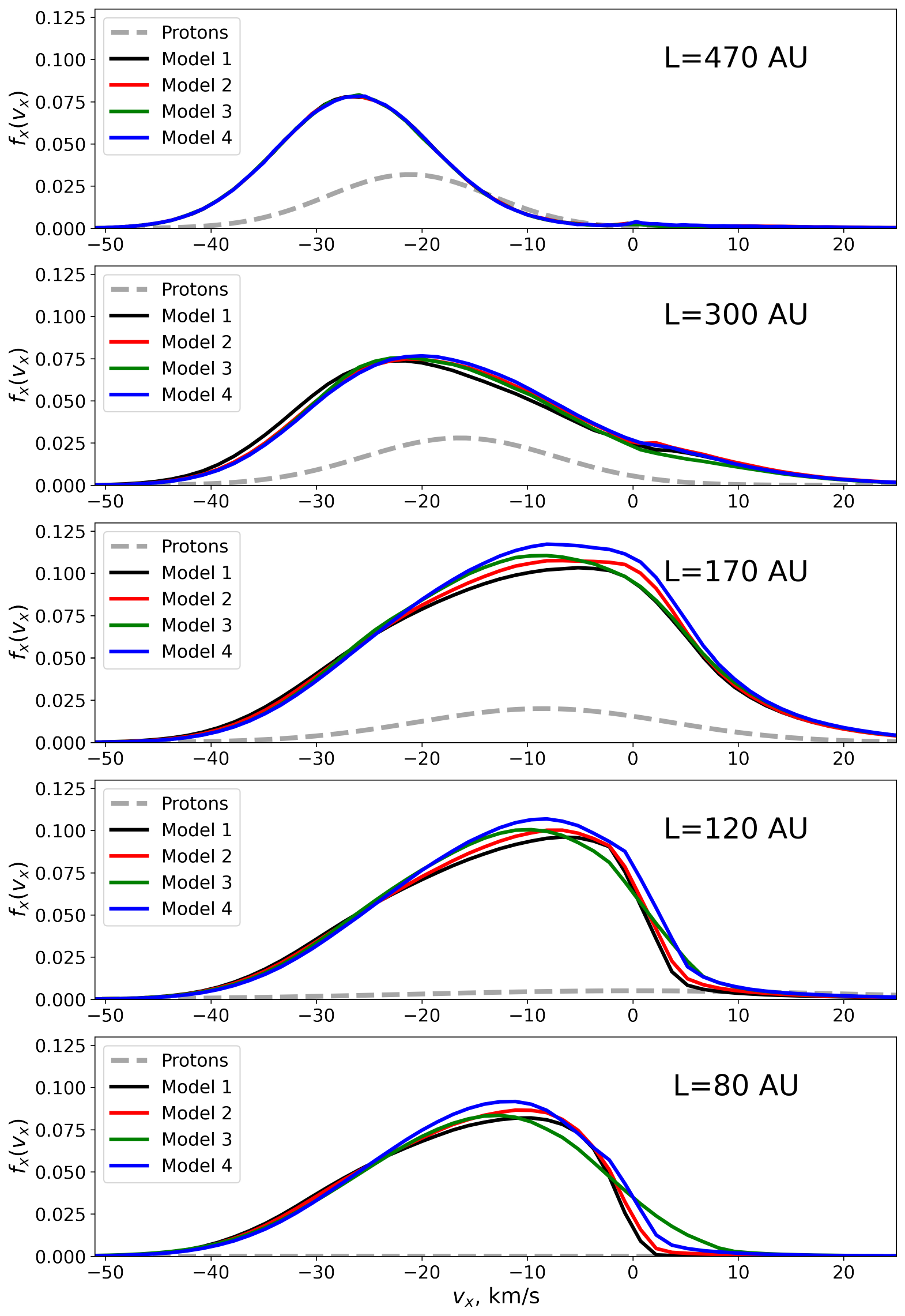}
    \caption{ The $v_{x}$-projection of the velocity distribution function in the outer heliosheath (470, 300, and 170 AU), near the heliopause (120 AU), and near the termination shock (80 AU)}
    \label{fig:f_el}
\end{figure}
\begin{figure*}
    \centering
    \includegraphics[width=1\linewidth]{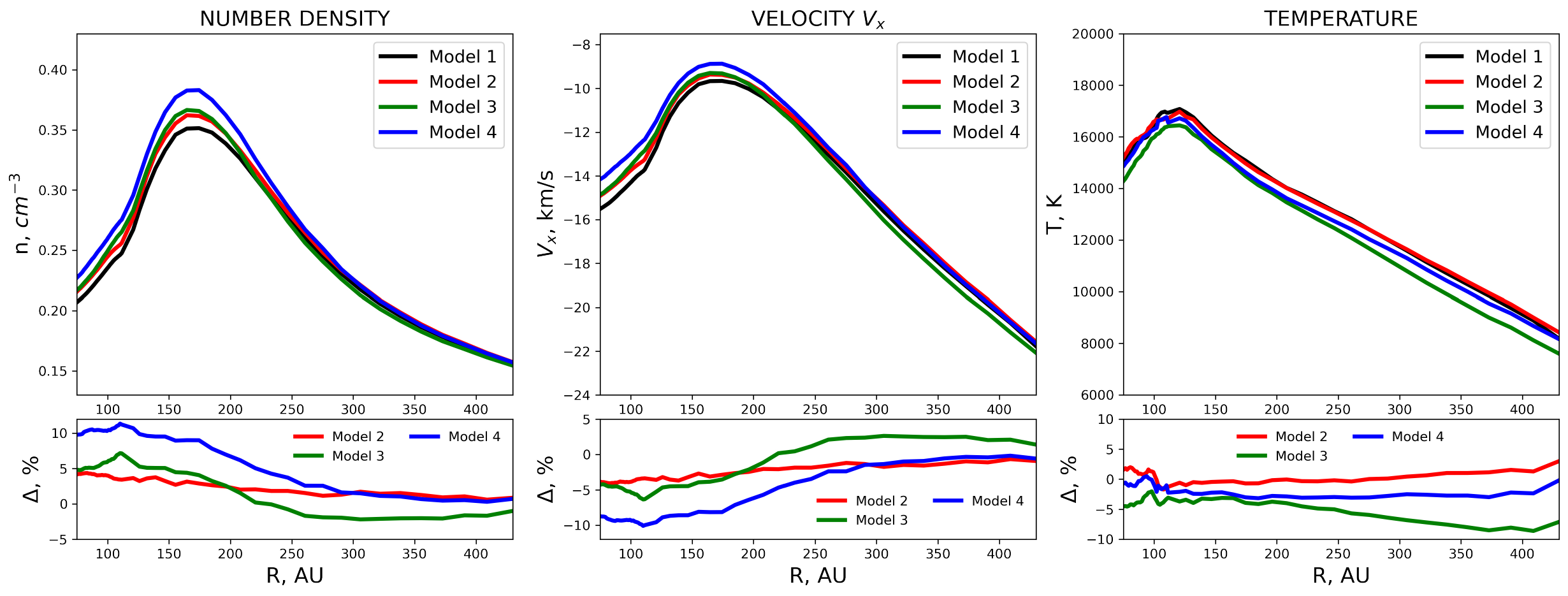}
    \caption{Number density (cm$^{-3}$), velocity (km/s) and temperature (K) as a function of heliocentric distance (AU). Lower panel shows the difference between the labeled model and Model 1 in \%}
    \label{fig:moments_el}
\end{figure*}
\begin{figure}
    \centering
    \includegraphics[width=1\linewidth]{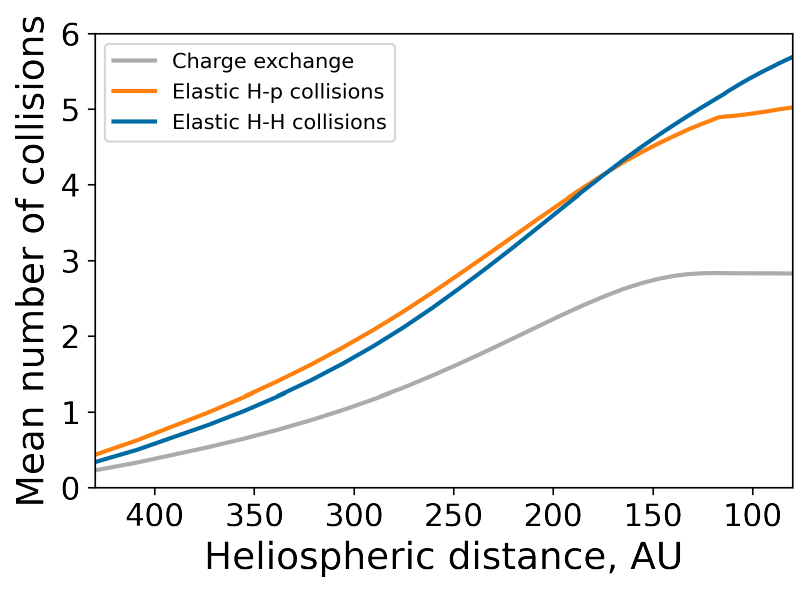}
    \caption{The mean number of collisions as a function of the heliocentric distance for each type of scattering}
    \label{fig:n_collisions}
\end{figure}
\section{Results} \label{sec: results}

\subsection{Effects of elastic H-p and H-H collisions}

Firstly, we explore the effects of elastic H-p and H-H collisions. To do this, we performed calculations in the frame of the following models: 
\begin{itemize}
    \item Model 1 is a kind of "standard" approach that is used in all modern models of the global heliosphere. Only the charge exchange process is taken into account in this approach. Also, no angular scattering during charge exchange is assumed, i.e. the interaction of particles does not change their velocity directions and absolute values, and the only electron jumps from the H atom to the proton. Under this approach the scattering angle $\chi$ is always equal to $\pi$. Hereafter, we refer to this as $\chi=\pi$  approximation. Section \ref{section: angular scattering} explores uncertainties introduced by this approximation. Note, also, that here we use the cross section \ref{sigma_tot_ex} found using the data by \cite{schultz_2016};  \\
   
    \item In Model 2 we consider combined effects of two processes: 1) charge exchange in  $\chi=\pi$  approximation and  2) elastic H-p collisions. This model explores how effects of elastic H-p collisions change the distributions of Model 1;\\
    
    \item In Model 3 we consider combined effects of two processes: 1) charge exchange in  $\chi=\pi$  approximation and  2) elastic H-H collisions. This model explores how effects of elastic H-H collisions change the distributions of Model 1;\\
    
    \item Model 4 is the most complete. It incorporates the effects of three processes: 1) charge exchange in $\chi=\pi$  approximation, 2) elastic H-p collisions, and 3) elastic H-H collisions. 

\end{itemize}
The results of these modeling calculations are presented in Figures \ref{fig:f_el} and \ref{fig:moments_el}. Figure \ref{fig:f_el}  presents  distribution functions $f(x, v_x)$ which are integrals of $f_H$ over $v_y$ and $v_z$:  $f(x, v_x)= \int f_H dv_y d v_z$. The functions $f(x, v_x)$ are shown at five different heliocentric distances. Figure \ref{fig:moments_el} presents moments of the velocity distribution (number density, bulk velocity and temperature) as functions of the heliocentric distance.

Before discussing the effects of elastic collisions we would like to remind of the basic effects of the charge exchange process (Model 1). 
For the scattering angle $\chi= \pi$ the velocity of the atom after collision is equal to the velocity of the proton before the collision. 
The bulk velocity of the proton component in the outer heliosheath (between the bow shock and heliopause) is smaller than the LISM velocity (see Figure \ref{figure: plasma_parameters}). The newly created (by charge exchange) atoms have smaller velocities on average as well. As a result, the maximum of the H atom velocity distribution moves to the smaller velocities (to the right in Figure \ref{fig:f_el}) as the heliocentric distance approaches the Sun. From 26 km/s at $\sim$500 AU (top panel in Figure \ref{fig:f_el}) it reduces to 20 km/s as far as the at 300 AU. At distances of 80-200 AU the maximum is about 10 km/s.

Since the temperature of the proton component increases toward the heliopause, the dispersion of the velocities of newly created atoms is larger than in LISM. It can be seen from Figure \ref{fig:f_el} that already at L= 300 AU the velocity distribution is visibly broader compared to the distribution at L = 470 AU. The closer to the Sun, the greater the width of the distribution function becomes. 

The large width of the distribution function together with the deceleration of the bulk velocity leads to the creation of particles with positive velocities which move out of the Sun back to the interstellar medium (see, for example, bottom and second bottom panel of Figure 2). These particles with positive radial velocities are still significant in the velocity distribution at L=300 AU. However, this component almost disappears (due to charge exchange, of course) at L=470 AU. As it can be seen in the top panel of Figure \ref{fig:f_el}, only a small bump at 0 km/s resembles the existence of the atoms with positive velocities.

Another noticeable feature in the distribution function which is more visible at L=120 AU is the asymmetry with respect to its maximum.
The asymmetry appears due to the so-called selection effect. This effect consists of a higher disappearance (due to charge exchange) of the atoms with slower velocities during their penetration inside the heliosheath.
This happens simply because slower atoms need more time to penetrate to the same distance compared to fast atoms. During this time they have more chance to be charge exchanged with protons. The effect of selection is discussed, for example, in \cite{izmodenov2001}. Closer to the Sun the bulk velocity of H atoms decreases, which is the reason why the asymmetry is more pronounced at 120 and 80 AU.  The proton bulk velocity in the inner heliosheath (between the heliopause and the termination shock) changes to positive values and increases drastically as well as the temperature. The atoms created in this region may have high positive velocities which do not contribute to the distribution at the presented in the Figure \ref{fig:f_el} velocity limits. Due to the selection effect more slow H atoms interact with protons creating new high velocity atoms. For this reason, at 80 AU the maximum of the distribution function at 80 AU seems a little bit lower than at 120 AU, as the number of slower H atoms decreases. 

The influence of the charge exchange process can also be observed in the moments of the distribution function (see Figure \ref{fig:moments_el}). After interaction with protons, the bulk velocity decreases by absolute value. This process begins at L > 400 AU and develops  approaching the heliopause, where the bulk velocity achieves approximately 10 km/s. Then, upon passing the heliopause, the H atoms accelerate slightly up to 15-16 km/s due to the selection effect discussed above. 

The deceleration of H atoms due to charge exchange leads to an increase in their number density, which peaks near the heliopause, where the protons are compressed the most. The region of higher number density of H atoms in the outer heliosheath is known as the hydrogen wall. A higher dispersion of proton velocities, which increases when approaching the heliopause, results in a rise in H atom temperature compared to LISM parameters.

The results obtained above may differ quantitatively from the other models due to the limitations of our one-dimensional model. Unlike 2D or 3D models, where atoms can escape the heliosphere due to their tangent velocity components and the limited size of the heliopause, the 1D model simplifies the heliopause to an infinite plane perpendicular to the x-axis. Consequently, all atoms with a velocity component $v_x$  directed toward the heliopause are constrained to cross it, which influences the calculated moments of the distribution. Despite these limitations, the primary objective of this study is to qualitatively demonstrate the effects of elastic collisions rather than to provide precise quantitative predictions.

Firstly, we will investigate the influence of H-p elastic collisions introduced in Model 2. In Figure \ref{fig:f_el}, at 470 AU, there are no visible differences between Model 1 and Model 2; however, at 300 AU, the maximum of the distribution function shifts to the smaller absolute velocities as compared to Model 1. These changes occur for the following reasons: H-p elastic collisions happen more frequently (see Figure \ref{fig:n_collisions}) than charge exchange due to their larger total cross-section (Figure \ref{figure: cs_comparison}).  At each elastic scattering the H atom velocity changes toward the small (absolute) values. Most of time the scattering happens at small scattering angles so in one act of scattering the change of velocity is really small. Nevertheless, large number of scattering leads to visible shift to the right of the velocity distribution for Model 2 as compared with Model 1 already at 300 AU. H-p elastic collisions also produce more low (absolute) velocity atoms which is demonstrated in \ref{section:test_problem} in the frame of a toy model of passing cold beam of H atoms through a layer of protons with a homogenic distribution.

 At 170 AU, the maximum of the distribution function is noticeably higher than in Model 1 and is also shifted toward positive velocities. At this distance, there is a region of maximum number density for both H atoms and protons. Therefore, the number of collisions  increases. Thus, the effects produced by these collisions are more pronounced compared to those at 470 and 300 AU. For similar reasons, at 120 and 80 AU, the differences in the distribution functions become less prominent as the number density of protons decreases and, in addition, relative atom-proton velocity increases that leads to weaker coupling of atoms and protons because the total and momentum cross-sections decrease with increasing relative velocity. Nevertheless, at these distances, the maximum of the distribution function remains higher than in Model 1.

The difference in moments of the velocity distribution is not as pronounced at 400-500 AU but increases as we approach the heliopause, reaching nearly 5\% for both bulk velocity and number density. As shown in Figure \ref{fig:moments_el}, H atom deceleration is more efficient in Model 2 than in Model 1, and the hydrogen wall also increases by a few percent. Changes in H atom temperature are less obvious; at some distances, it is higher than in Model 1, although near the hydrogen wall, H atoms become cooler. Overall, these differences are less noticeable in percentage terms compared to those for number density and bulk velocity.

Next, we explore the effects of H-H elastic collisions (Model 3). Similar to Model 2, at 470 AU, there are no distinguishable differences in the distribution function. At 300 AU, the maximum of the distribution function is slightly higher and shifted toward positive velocities. Major changes occur closer to the Sun: the asymmetry of the distribution function noticeably decreases compared to Model 1. As H atoms interact with each other, they exchange momentum and energy, leading to a redistribution of velocities. This effect is the most pronounced at 80 AU, where the asymmetry nearly disappears. Thus, with H-H elastic collisions taken into account, the distribution function tends to maxwellian. 

Model 1 and 3 differences in the number density and bulk velocity are nonlinear: at large heliocentric distances, H-H collisions accelerate the H atoms by a few percent while lowering the number density. However, in the vicinity of the hydrogen wall, the atoms decelerate, and the number density increases almost at the same rate as in Model 2. Conversely, the H atom temperature is consistently lower than in Model 1 (by up to 5-10\%). H-H collisions reduce the distinction between the interstellar component and atoms created through charge exchange, which may contribute to this decrease in temperature.

Although both the total and momentum cross sections of H-H collisions are significantly lower than those of charge exchange, the number density of H atoms is much higher than that of protons. Since the number of collisions depends on both the cross section and number density, the number of H-H collisions is still higher than that of charge exchange (Figure \ref{fig:n_collisions}). As a result, the influence of H-H collisions is noticeable despite the lower total cross section. 

Finally, Model 4 includes both types of elastic collisions. The distribution function of Model 4 differs the most from that of Model 1: the shift and growth of the maximum are more pronounced (compared to the previous models), especially in the hydrogen wall (170 AU) and in the inner heliosheath (120 and 180 AU). Moreover, H-H collisions slightly decrease the asymmetry of the distribution function, particularly closer to the Sun, although not as significantly as in Model 3. 

From 500 to 300 AU, the bulk velocity and number density do not differ much from those in Model 2. However, in the vicinity of the hydrogen wall, H-H collisions amplify the deceleration effect produced by H-p collisions, leading to a 10\% decrease in bulk velocity and a corresponding increase in number density. The temperature curve of Model 4 lies between those of Model 2 and Model 3. Overall, the temperature of H atoms in Model 4 is lower than that in Model 1, but only by a few percent.

Overall, elastic collisions, particularly H-p collisions, significantly influence the distribution by further decelerating hydrogen atoms and increasing the density of the hydrogen wall. H-H collisions, meanwhile, reduce the asymmetry of the velocity distribution, bringing it closer to a maxwellian shape, particularly near the Sun. They also cause minor changes in bulk velocity and number density, with a slight acceleration at larger distances and deceleration near the hydrogen wall. The combined effect of elastic collisions, as demonstrated in Model 4, leads to a greater reduction in bulk velocity and a denser hydrogen wall than when only charge exchange is considered.
\begin{figure}
    \centering
    \includegraphics[width=1\linewidth]{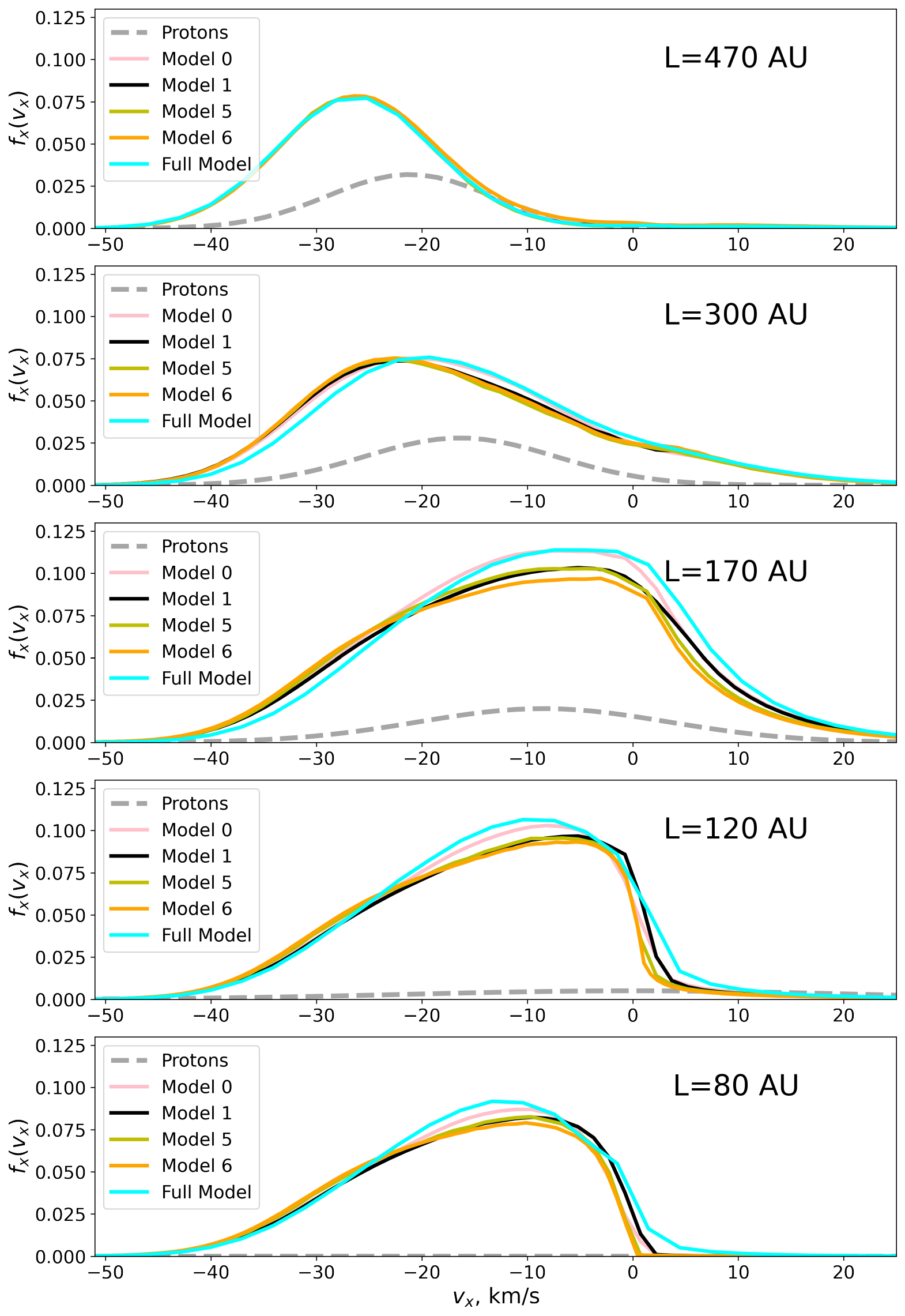}
    \caption{The $v_{x}$-projection of the velocity distribution function in the outer heliosheath (470, 300, and 170 AU), near the heliopause (120 AU), and near the termination shock (80 AU)}
    \label{fig:f_ex}
\end{figure}
\begin{figure*}
    \centering
    \includegraphics[width=1\linewidth]{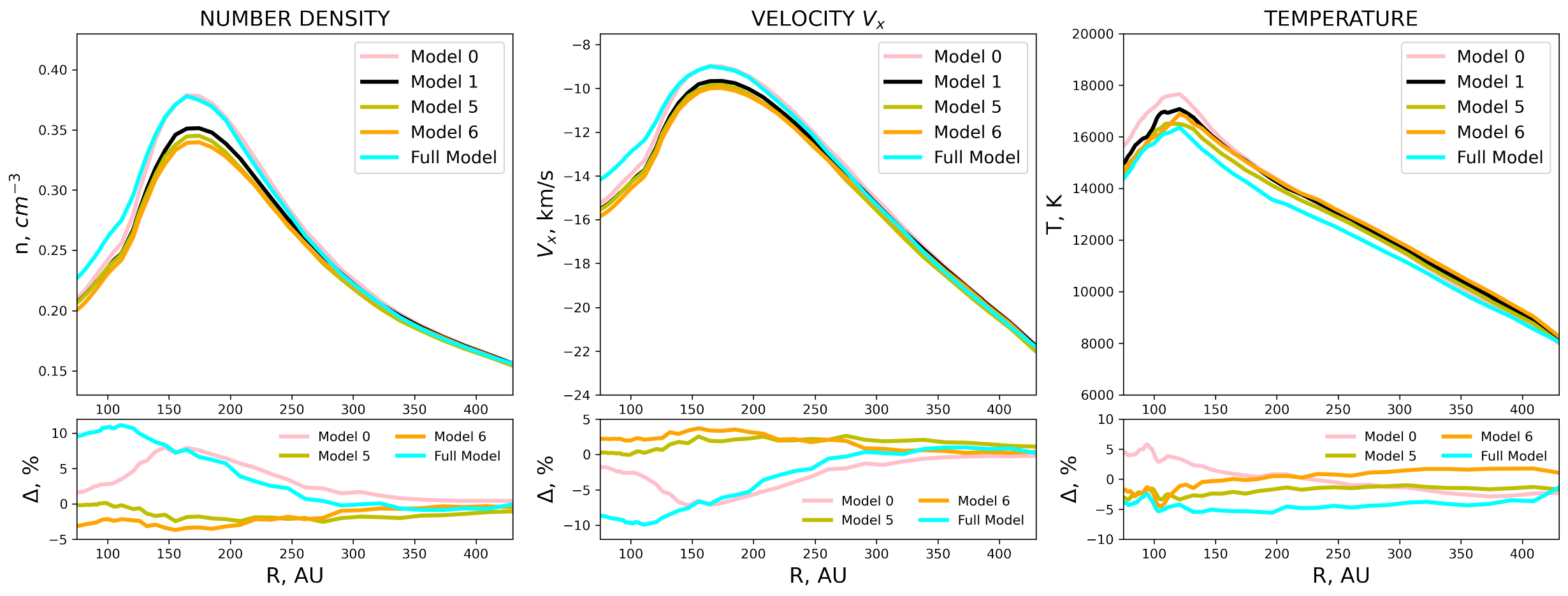}
    \caption{Number density (cm$^{-3}$), velocity (km/s) and temperature (K) as a function of heliocentric distance (AU). Lower panel shows the difference between the labeled model and Model 1 in \%}
    \label{fig:moments_ex}
\end{figure*}

\subsection{Angular scattering charge exchange} \label{section: angular scattering}

In this subsection we explore effects of angular scattering charge exchange. For the discussion on cross section see \ref{appendix: cross section}. We performed calculations in the frame of the following models: 
\begin{itemize}
    \item Model 0 that includes charge exchange with $\chi=\pi$ (where $\chi$ is the scattering angle) approximation and \cite{lindsay_2005} cross section. This approach is used in the global modeling by \cite{izmodenov_2020} and later works. The total cross section by \cite{lindsay_2005} is a fit of the experimental data and its values are higher than those by \cite{schultz_2016} especially at low relative velocities. Also note that the \cite{lindsay_2005} cross section can be used only for a specific range of velocities (see \ref{appendix: cross section}); \\
    
    \item Model 1 that includes charge exchange with $\chi=\pi$  approximation and the fit (\ref{sigma_tot_ex}) of the total charge exchange cross section presented in \ref{appendix: cross section} as in previous paragraph; \\
    
    \item  Model 5 that includes angular scattering charge exchange with the differential cross section presented in \ref{appendix: cross section}. This model explores how angular dependence changes the distributions of Model 1; \\
    
    \item Model 6 is an attempt to make a simple approximation for future global modeling that would include the effects of Model 5. To do that we took the fit of the momentum transfer cross section presented in \ref{appendix: cross section} and assumed that the scattering angle $\chi=\pi$. In this case,  $\sigma_{mt, ex} = 2 \sigma_{tot,ex}$, thus, the total cross section can be found. \\
    
    \item Full Model takes into account both H-p and H-H elastic collisions in addition to charge exchange. Moreover,  the angular scattering charge exchange with the differential cross section presented in \ref{appendix: cross section} is considered. This model considers all the effects discussed in this paper.
\end{itemize}

Results of the modeling calculations are presented in Figures \ref{fig:f_ex} and \ref{fig:moments_ex}. Figure \ref{fig:f_ex} presents distribution functions $f(x, v_x) = \int f_H dv_y d v_z$ at different heliocentric distances. Figure \ref{fig:moments_ex} presents moments of the velocity distribution - number density, bulk velocity and temperature  - as functions of the heliocentric distance. 

Firstly, we discuss results of Model 0. The charge exchange cross-section used in this model is higher than the theoretical cross-section of  \cite{schultz_2016} at the range from 1 to 100 km/s, leading to an increased number of collisions. At 470 AU, Model 0 and Model 1 show minimal differences. However, at 300 AU, the distinctions become more pronounced. Model 0 demonstrates a more noticeable shift in the peak toward lower velocities compared to Model 1, indicating that the charge exchange in Model 0 operates more "effectively" due to the higher total cross-section.

The difference between the two models becomes substantial at 170 AU, where the peak in Model 0 is significantly higher. At 120 AU and 80 AU, Model 0 continues to show a higher velocity peak, and the distribution function in Model 1 appears more asymmetric in comparison.

In terms of number density, Model 0 exhibits a higher peak than Model 1. Between 100 and 150 AU, Model 0 shows a steeper rise in density, while Model 1 increases more gradually. After the peak, both models experience a decline in number density, though Model 0 remains consistently higher. Model 0 also predicts a slightly lower bulk velocity compared to Model 1, as the number of H atoms interacting with protons increases. Both models exhibit a velocity peak around 150 AU, but Model 0 has a sharper peak.

The primary differences in both number density and bulk velocity are observed near the hydrogen wall, where deviations between the two models approach 7\%. At larger distances and closer to the Sun, the differences are smaller, around 2-3\%. Model 0 predicts a higher temperature than Model 1 throughout the entire range of distances. Model 0 consistently remains hotter, with the most significant difference occurring between 80 and 150 AU, where it is approximately 5\% hotter. Overall, the larger cross-section of the charge exchange rate (in $\chi = \pi$ approximation) leads to the effects similar to elastic collisions for number density and bulk velocity, but the opposite effect for the temperature.

At first glance, the Model 5 distribution does not differ significantly from that of Model 1; however, there are still effects worth discussing. Already at 300 AU, Model 5 exhibits a slightly lower peak, shifted toward higher velocities compared to Model 1. These changes become more pronounced at closer heliospheric distances: the major differences are observed near $v_x=0$, where the right tail of the Model 5 distribution shifts toward negative values. Due to the scattering angle of charge exchange not being exactly 180$^{\circ}$, newly formed atoms do not acquire the exact velocity of their parent protons. Consequently, momentum transfer is reduced, leading to smaller modification of the distribution function.

The differences in moments of the velocity distribution are also not particularly prominent. Model 5 predicts a higher bulk velocity and a slightly lower number density than Model 1 near the hydrogen wall and throughout most regions. In the vicinity of the hydrogen wall, the deviations reach approximately 2\% but nearly vanish after passing the heliopause (120-100 AU). This may be related to the fact that as relative velocity increases, the momentum transfer cross-section for angular charge exchange approaches the $\chi=\pi$ approximation, and at high velocities, they nearly coincide. In terms of temperature, Model 5 predicts a slightly lower temperature than Model 1 at all distances. The temperature difference becomes more evident between 80 and 150 AU, where Model 1 temperature consistently remains higher than Model 5 one. Due to the angular dependence, the velocity dispersion of H atoms decreases. When $\chi=\pi$, the atom acquires the exact velocity of the parent proton. As the proton bulk velocity is lower by absolute value the newly created atoms will make the distribution wider. With angular scattering, the velocity of the scattered atom lies between that of the parent atom and the proton, but closer to the proton, as the peak of the differential cross section is near 180. As a result, the velocity dispersion becomes smaller.

The inclusion of angular charge exchange into the global model is a complex problem. Treating this problem directly may significantly increase computational time. Therefore, we decided to create a simpler approximation that considers the effects of angular transfer and can be easily applied. We propose that the primary difference between angular transfer and the $\chi=\pi$ approximation lies in the lower rate of momentum transfer. Thus, if we use the momentum transfer cross-section and assume the $\chi=\pi$ approximation, we will obtain a total cross-section that takes into account the influence of angular dependence. Model 6 is calculated using this new cross-section. Although the distribution functions tend toward those of Model 5, there is no exact match. At 170 and 80 AU, the maximum is noticeably lower. More differences are observed in the moments: H atoms decelerate slightly more in Model 6 than in Model 5. The most visible difference lies in the temperature of the H atoms, where results of Model 6 are close to Model 1. The smaller temperature obtained in Model 5 is due to angular scattering as it has been discussed above.

Finally, we present the most complete model where angular charge exchange and elastic H-p and H-H collisions are considered. The angular charge transfer accelerates atoms slightly compared to the Model 3 with $\chi=\pi$ approximation, however, the effects of H-p and H-H elastic collisions described in the previous section are those that determined the main difference with Model 1. Moreover, the temperature decreases even more than in Model 3. Overall, compared to the charge exchange with $\chi=\pi$ approximation (Model 1) the differences in the bulk velocity and number density
are up to 10\% and the temperature declines up to 5\%.

\subsection{Observable parameters}
In this paper, we calculate the line of sight absorption profile by the heliospheric interface along the 1D line toward the interstellar flow direction. Since the source Lyman-$\alpha$ profile is not known, we calculate the relative absorption $I/I_0$. To model the spectra we follow the procedure provided by \cite{izmodenov2002}, where the profile along the line of sight is calculated as:
\[
I(\lambda)=I_{0}(\lambda) \exp \left(-\frac{\pi e^{2} f N \lambda^{2}}{m c^{2}} \phi(v_{los})\right),
\]
where $I_{0}$ is the assumed background Ly-$\alpha$ profile, $f$ is the oscillator strength, $N$ is the column density, $\lambda$ is the wavelength, and $\phi(v_{los})$ is the normalized velocity distribution along the line of sight. Due to the Doppler effect the wavelengths can be transferred to the line-of-sight velocity (the projection of velocity on the line of sight): $v_{los}=c\left(1-\lambda_0 / \lambda\right)$, where $\lambda_0$ is the Lyman-$\alpha$ line that equals 1215.66 Å.
\begin{figure}[t]
\center{\includegraphics[width=1\linewidth]{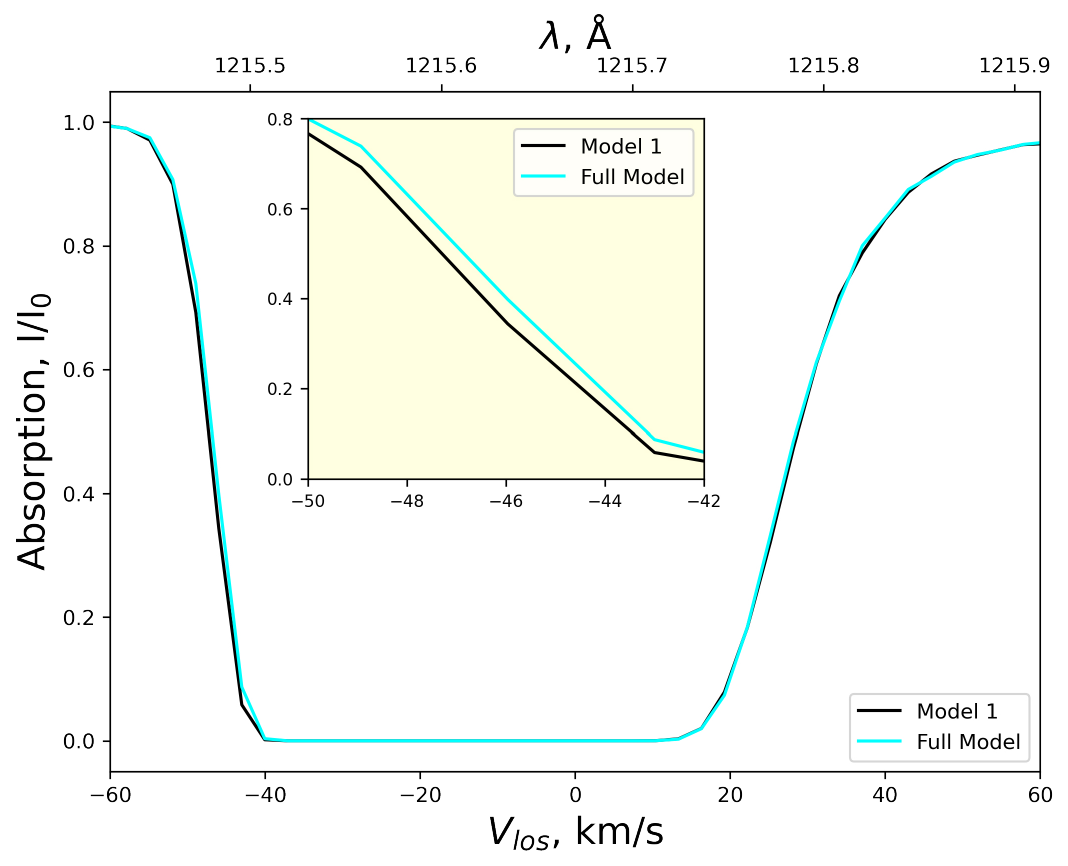}}
    \caption{Ly-$\alpha$ absorption by layer along the line of sight in upwind direction for different models}
    \label{figure:ly_a_abs}
\end{figure}
\begin{figure}[t]
\center{\includegraphics[width=1\linewidth]{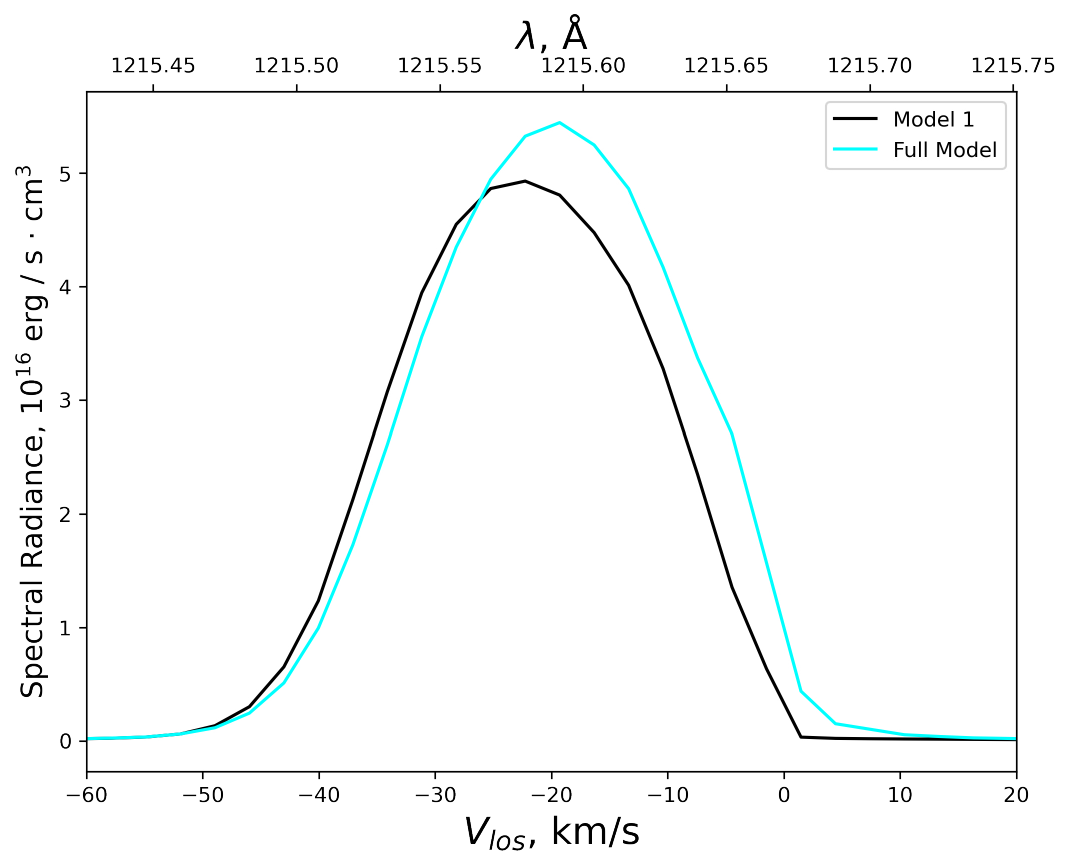}}
    \caption{Backscattered solar Ly-$\alpha$ emission at 2 AU along the line of sight in upwind direction for different models}
    \label{figure: ly_a_back}
\end{figure}

Figure \ref{figure:ly_a_abs} presents a comparison of Lyman-$\alpha$ absorption profiles as a function of line-of-sight velocity $v_{los}$  for two models: Model 1 and the Full Model. The bottom x-axis represents the velocity $v_{los}$ in km/s, the corresponding wavelengths are shown on the top. The y-axis shows the relative absorption $I/I_0$. Both models exhibit similar overall absorption behavior. The curves are hardly distinguishable, however, the left tail of the Full Model spectra is slightly shifted to the positive velocities. It can be seen more precisely on the zoomed in inset: the difference is $\sim$1 km/s. Thus, it can be seen that H-p and H-H elastic collisions have almost no influence on the absorption spectra.

We also computed the backscattered solar Lyman-$\alpha$ profiles along the line of sight at 2 AU in the upwind direction using the method presented in \cite{KATUSHKINA2011}. The line of sight in the calculations is directed toward the interstellar flow as for the absorption profile. The calculations were performed using a self-absorption approach. In this method, assuming a radial line-of-sight, the solution to the transfer equation (Eq. (1) in \cite{KATUSHKINA2011}) can be expressed as:
\begin{equation*}
\begin{aligned}
& I(\mathbf{r}, \lambda, \boldsymbol{\Omega}) = \lambda_{0} \sigma_{t o t} F_{S, 0} s_{0}\left(2 \lambda_{0}-\lambda\right) r_{E}^{2} \frac{11 / 12+1 / 4}{4 \pi} \\
& \cdot \int_{0}^{\infty} \frac{f_H(\mathbf{r}+s \boldsymbol{\Omega}, v_{los})}{|\mathbf{r}+s \boldsymbol{\Omega}|^{2}} e^{-\tau_{\lambda}(\mathbf{r}+s \boldsymbol{\Omega}, \mathbf{r})} d s,
\end{aligned}
\end{equation*}
where $\mathbf{r}$ is an observer's position,  $\boldsymbol{\Omega}$ is a line-of-sight direction, s is a coordinate along the line-of-sight, $\tau_{\lambda}(\mathbf{r}+s \boldsymbol{\Omega}, \mathbf{r})$ is the optical thickness for scattered photons with the wavelength $\lambda$ calculated from the scattering point $\mathbf{r'} = \mathbf{r} + s\boldsymbol{\Omega}$ to the observer located at the point $\mathbf{r}$, $F_{S,0} \approx 3.32\cdot 10^{11}$ s$^{-1}$cm$^{-2}$  is the flux at 1 AU , $s_{0}\left(2 \lambda_{0}-\lambda\right)$ -- the normalized solar Ly-$\alpha$ profile at 1 AU, and $\sigma_{tot}= \frac{\pi e^{2}}{m c} f$.

The results are shown in Figure \ref{figure: ly_a_back}. The differences between the models are more pronounced than in the case of the absorption spectra. The peak of the Full Model profile is significantly higher and shifted toward positive velocities by approximately 2 km/s compared to Model 1. These profile changes reflect those of the hydrogen atom distribution function, where the maximum was also shifted to smaller (in absolute value) velocities.

\begin{table}
\resizebox{\columnwidth}{!}{%
\begin{tabular}{l|l|l|l|}
\cline{2-4}
                                       & Intensity, R   & LOS Velocity, km/s & LOS Temperature, K \\ \hline
\multicolumn{1}{|l|}{Model 1}  & 712        & -22.4            & 13495           \\ \hline
\multicolumn{1}{|l|}{Full Model}      & 799 (+12\%) & -20.0 (-11\%)     & 14533 (+8\%)      \\ \hline

\end{tabular}%
}
\caption{Spectral Ly-$\alpha$ moments for different models}
\label{tab:my-table}
\end{table}
\begin{figure}[t]
\center{\includegraphics[width=1\linewidth]{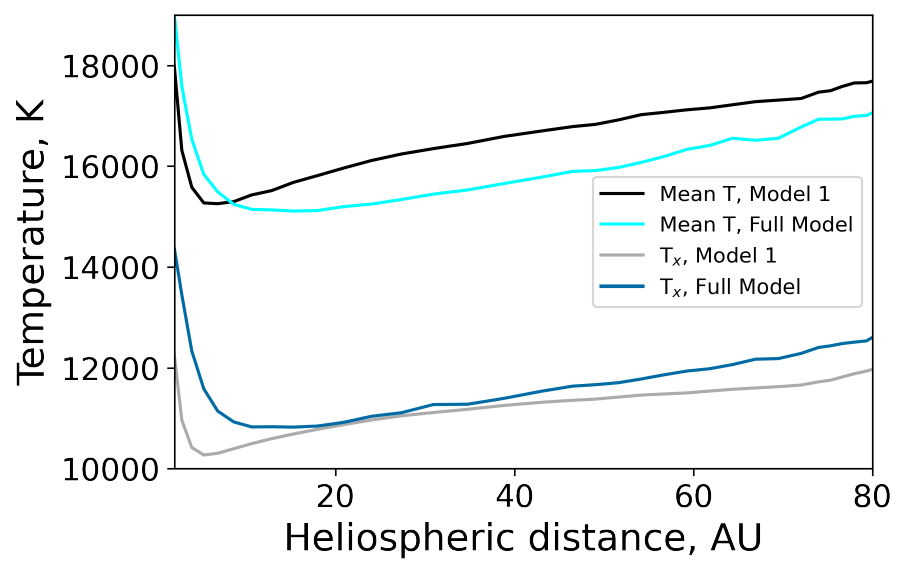}
    \caption{Comparison of mean ($T=1/3 \cdot (T_x+T_y+T_z)$) temperature and $T_x$ for Model 1 and Full Model inside the heliosphere}
    \label{figure:t_x}}
\end{figure}

Knowing the Ly-$\alpha$ profile, it is possible to calculate its moments: 
\begin{gather*}
I_{\text {los }}(\mathbf{r}, \boldsymbol{\Omega})=\frac{4 \pi}{10^{6}} \int_{0}^{\infty} I(\mathbf{r}, \lambda, \boldsymbol{\Omega}) d \lambda \quad  \text {(intensity in rayleighs) } \\
V_{\text {los }}(\mathbf{r}, \boldsymbol{\Omega})=\frac{\int_{0}^{\infty} v_{los}(\lambda) I(\mathbf{r}, \lambda, \boldsymbol{\Omega}) d \lambda}{I_{\text {los }}(\mathbf{r}, \boldsymbol{\Omega})} \\
\text {(Doppler shift or line-of-sight average velocity in km/s) } \\
T_{\text {los }}(\mathbf{r}, \boldsymbol{\Omega})=\frac{m_{H}}{k_{b}} \frac{\int_{0}^{\infty}\left(v_{los}(\lambda)-V_{\text {los }}(\mathbf{r}, \boldsymbol{\Omega})\right)^{2} I(\mathbf{r}, \lambda, \boldsymbol{\Omega}) d \lambda}{I_{\text {los }}(\mathbf{r}, \boldsymbol{\Omega})} \\
\text{(line width or line-of-sight temperature in K)}
\end{gather*}

Table \ref{tab:my-table} shows the moments of distributions (intensities, line-of-sight (LOS) velocities, line-of-sight (LOS) temperatures) for two models. Full Model shows a 12\% increase in intensity, an 11\% reduction by absolute value in LOS velocity, and an 8\% increase in LOS temperature compared to Model 1. The changes in intensity and LOS velocity can be directly attributed to the hydrogen atom moments: the number density and average velocity, with elastic collisions and angular transfer taken into account, change at almost the same rate. On the other hand, while the LOS temperature increases, the hydrogen atom temperature actually decreases in comparison to Model 1.  In the previous paragraphs, we calculated the mean temperature $T=\frac{1}{3}(T_x +T_y+T_z)$. The LOS temperature is connected to the radial kinetic temperature and when an observer looks toward upwind it equals $T_{x}$. The temperature components differ from each other, and the Full Model $T_{x}$ at lower heliocentric distances becomes slightly higher than in Model 1 (see Figure \ref{figure:t_x}).

\section{Conclusions} \label{section: conclusions}

In this work, we studied the effects produced by elastic H-P and H-H collisions and by the angular scattering during charge exchange on the velocity distribution of the interstellar H atoms in the heliosphere and on the moments of the velocity distribution. The effects are visible but not large: 

1. H-p collisions lead to formation of more particles with smaller velocities making the velocity distribution function wider. This leads to effective deceleration (reducing bulk velocity) and "heating" (i.e. increasing the kinetic temperature) of the H atom population

2. H-H collisions reduce the asymmetry of the velocity distribution, particularly in regions closer to the Sun, and drive the distribution toward a more maxwellian shape. Additionally, H-H collisions slightly lower the temperature of hydrogen atoms compared to the model without elastic collisions.

3. The angular scattering leads to a decrease in the temperature and a slight acceleration of hydrogen atoms compared to "$\chi = \pi$" approximation, although the effect of  angular scattering in the charge exchange is smaller than the effect of elastic collisions.

4. The net effect is as follows. The number density of H atoms inside the heliosphere is increased by 10 \% as compared with the "standard" model in which only charge exchange in the $\chi =\pi$ approximation is considered. The bulk velocity becomes smaller by $\sim$1.5 km/s that is about 8 \%. The "temperature" is decreased by $\sim$4-5 \% that is less than 1000 K.

In considered model the parameters of plasma component are fixed, so we did not consider how the considered processes would influence the plasma distribution. With the presented results the influence is expected to be small or even negligible. 

In addition, we explore how the considered effects influence such observables as solar backscattered  Ly-$\alpha$ spectra as well as the Ly-$\alpha$ absorption spectra produced by interstellar atoms in the heliosheath. The comparison of Lyman-$\alpha$ absorption spectra between Model 1 and the Full Model shows very similar overall behavior. The absorption profiles are nearly indistinguishable, with only a slight shift of the left tail in the Full Model by about 1 km/s. Thus, the inclusion of elastic collisions does not notably affect the Lyman-$\alpha$ absorption properties. In contrast, the backscattered Lyman-$\alpha$ profiles show more pronounced differences between the models. The Full Model displays a peak that is both higher and shifted towards positive velocities by approximately 2 km/s compared to Model 1. When examining the moments of the Lyman-$\alpha$ profiles (intensity, line-of-sight (LOS) velocity, and LOS temperature), the Full Model demonstrates a 12\% increase in intensity, an 11\% reduction in LOS velocity (by absolute value), and an 8\% increase in LOS temperature compared to Model 1.

\begin{acknowledgement}
The research is carried out using the equipment of the shared research facilities of HPC Resources of the Higher School of Economics (\cite{Kostenetskiy_2021}).
\end{acknowledgement}

\paragraph{Competing Interests}

None

\paragraph{Data Availability Statement}

The data underlying this article
will be shared on reasonable request to the corresponding author.

\printendnotes

\bibliography{refs}
\appendix
\section{Cross sections} \label{appendix: cross section}

\begin{figure}
    \center{\includegraphics[width=1\linewidth]{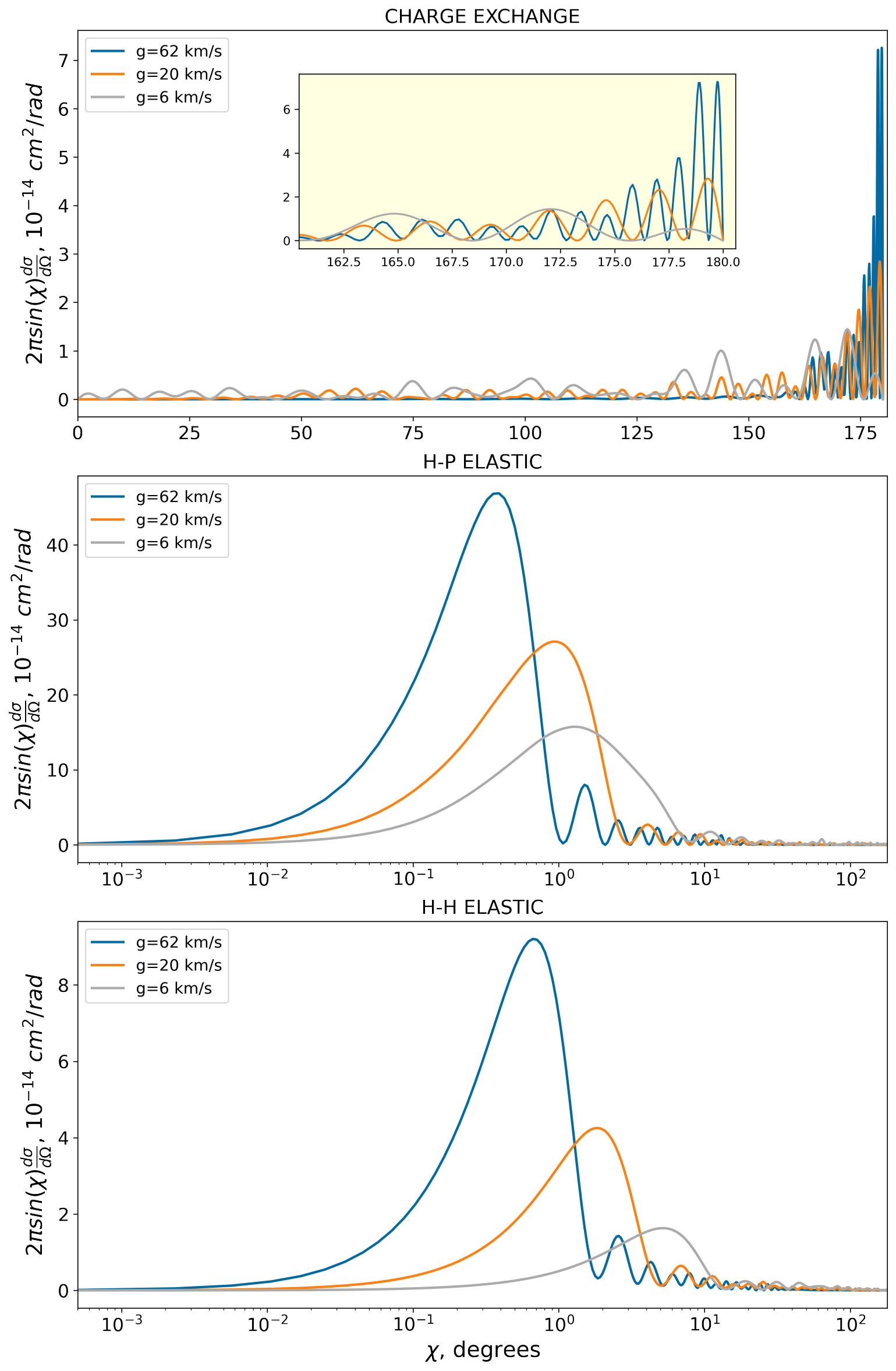}}
    \caption{
    Differential cross section for charge exchange, H-p and H-H elastic collisions at various relative velocities $g$}
    \label{figure: diff_cs}
\end{figure}
\begin{figure}
   \center{\includegraphics[width=1\linewidth]{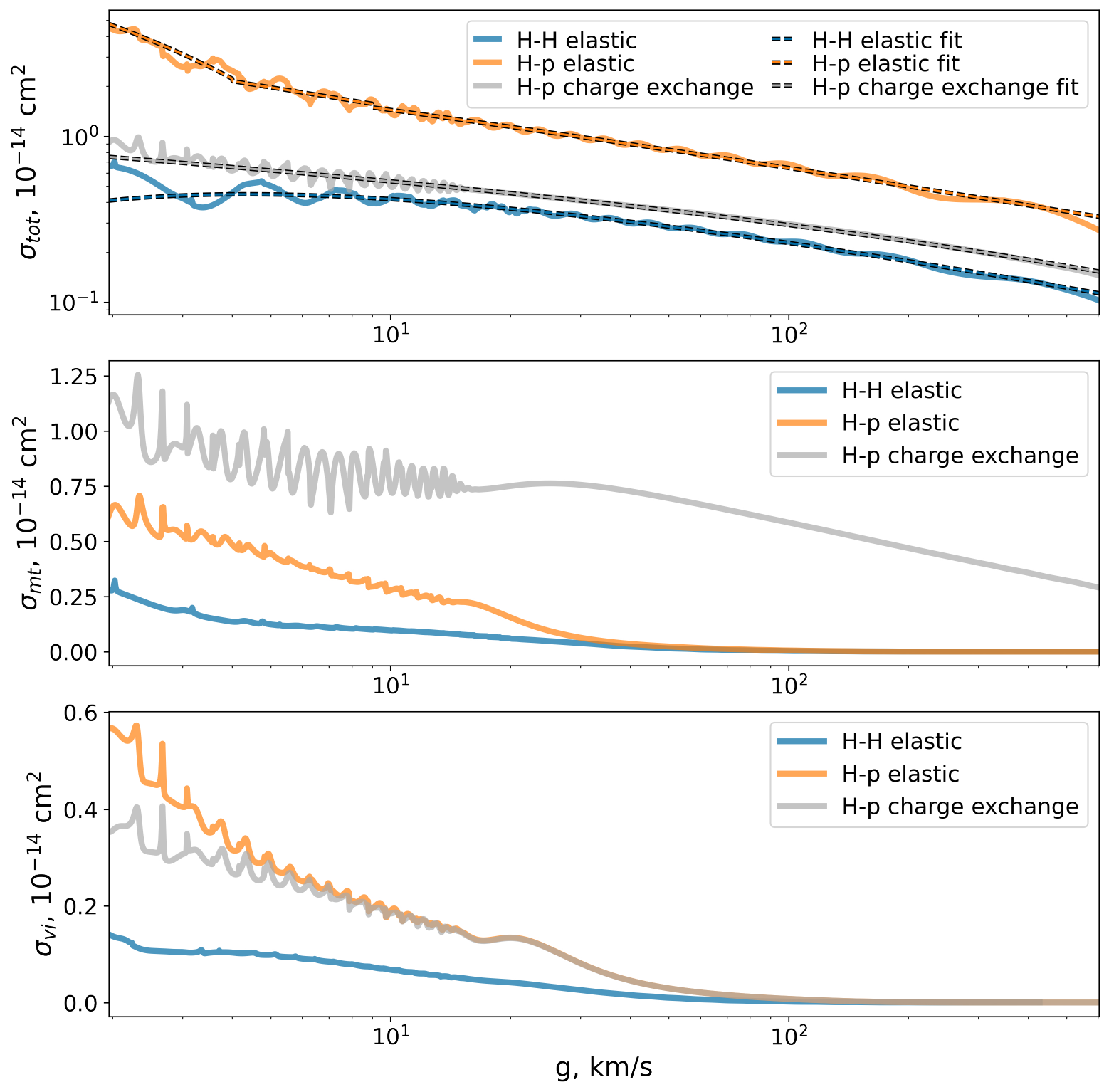}}
    \caption{
    Comparison of total (upper panel), momentum transfer (middle panel) and viscosity (lower panel) cross sections for charge exchange, elastic H-p scattering and elastic H-H scattering cross sections}
    \label{figure: cs_comparison}
\end{figure}
In this Appendix, we describe cross sections that have been employed in this work. 

For the charge exchange and elastic scattering we used the cross sections obtained by Schultz et al. (\citeyear{schultz_2016, schultz2023}). The cross sections have been calculated in the distinguishable particle approach.  In this approach the cross sections for the charge exchange and elastic collisions can be separated. This perfectly fits our purposes. The distinguishable approach works for high relative velocities ($g >3$ km/s). At around 0.2 km/s, the difference between indistinguishable and the sum of distinguishable charge exchange and elastic cross sections is approximately 40\%. (see \cite{schultz_2016} section 3.3). This discrepancy decreases to 5\% at approximately 2 km/s and becomes negligible beyond 196 km/s. The tests performed by \cite{schultz_2016} show that the error resulting from the distinguishable approach is insignificant at higher velocities, amounting to only 0.1\% at $g\approx20$ km/s, and up to 5\% at 2 km/s.  The mean relative velocity in the outer heliosheath is $\sim10$ km/s, therefore, this approach can be used.

The differential cross sections $d\sigma/d \Omega$ were downloaded from the following website: \url{https://sites.physast.uga.edu/amdbs/elastic/index.html}.
Examples of the differential cross sections for three values of the relative velocities are shown in Figure \ref{figure: diff_cs}  as a function of the scattering angle.
The elastic collision cross section scatters the direction of colliding particles by less than few degrees.
Conversely, the charge exchange cross section has a strong maximum at $\chi = \pi$ (Figure \ref{figure: diff_cs}, upper panel) where the maximum momentum exchange between colliding particles appeared.

Figure \ref{figure: cs_comparison} shows the total, momentum transfer and viscosity cross sections are the integrals of the differential cross sections:
\begin{equation}\label{sigma_tot}
\sigma_{tot}  = 2 \pi \int \frac{d\sigma}{d \Omega} sin \chi d \chi
\end{equation}
\begin{equation}\label{sigma_mt}
\sigma_{mt}  = 2 \pi \int \frac{d\sigma}{d \Omega} (1- cos \chi) sin \chi d \chi
\end{equation}
\begin{equation}\label{sigma_vi}
\sigma_{vi}  = 2 \pi \int \frac{d\sigma}{d \Omega} sin^3 \chi d \chi
\end{equation}
The cross sections are presented as a functions of the relative velocity of colliding particles.

To make the cross sections more suitable for modeling, we fit the total cross sections (see, also, Figure \ref{figure: cs_comparison} dashed curves) as follows: 

1. Charge exchange:
\begin{equation}\label{sigma_tot_ex}
    \sigma_{tot,ex} = (a_{1}-a_{2}ln(g))^2 \text{ ($g$ in cm/s),} 
\end{equation}
where $a_{1} = 1.87618291\cdot 10^{-7}$, $a_{2} = 8.28949260\cdot 10^{-9}$, and $g$ is the relative velocity.

2. H-p elastic collisions:

\begin{equation}
\sigma_{tot,Hp} = \left\{\begin{array}{l}
  (a_{1}-a_{2}ln(g))^2, \quad g<4\text{ km/s}\\
  (a_{3}-a_{4}ln(g))^2\cdot \left(1-\exp\left(\frac{-a_{5}}{g}\right)\right)^{a_6}, \quad 4\leq g<9\text{ km/s}\\
 (a_{7}-a_{8}ln(g))^2\cdot \left(1-\exp\left(\frac{-a_{9}}{g}\right)\right)^{a_10}, \quad g\geq15\text{ km/s}
\end{array}\right. 
\end{equation}
where $a_{1}=1.39317333\cdot 10^{-6}$,  $a_{2}=9.64544927\cdot 10^{-8}$,  $a_{3}=1.82272486\cdot 10^{-6}$, $a_{4}=1.77433130\cdot 10^{-9}$, $a_{5}=9.03662968\cdot 10^{-1}$, $a_{6}=3.89732769\cdot 10^{-1}$, $a_{7}=1.42076282\cdot 10^{-6}$, $a_{8}=4.24536719\cdot 10^{-7}$,  $a_{9}= 9.86302853\cdot 10^{-1}$,  $a_{10}=5.22372476\cdot 10^{-1}$.

3. H-H elastic collisions
\begin{equation}
    \sigma_{tot,HH} =  (a_{1}-a_{2}ln(g))^2\cdot \left(1-\exp\left(\frac{-a_{3}}{v}\right)\right)^{a_{4}}\text{ ($g$ in cm/s),}
\end{equation}
where $a_{1} = 1.77109098\cdot 10^{-5}$, $a_{2} = 1.76791997\cdot 10^{-6}$, $a_{3}=1.07425517$, $a_{4}=6.75707260\cdot 10^{-1}$

4. Model 6 charge exchange:

Here, we multiplied the momentum transfer cross section by 0.5 and fit the received curve:
\begin{equation}
\sigma_{tot,Hp} = \left\{\begin{array}{l}
  (a_{1}-a_{2}ln(g))^2\cdot \left(1-\exp\left(\frac{-a_{3}}{g}\right)\right)^{a_4}, \quad  g\leq 16\text{ km/s}\\
  (a_{11}-a_{12}ln(g))^2\cdot \left(1-\exp\left(\frac{-a_{13}}{g}\right)\right)^{a_{14}}, \quad  g>16\text{ km/s}\\
\end{array}\right. 
\end{equation}
where $a_{1}=7.16911227\cdot 10^{-8}$,  $a_{2}=4.27710011\cdot 10^{-8}$,  $a_{3}=8.90798298\cdot 10^{-1}$, $a_{4}=3.01412164\cdot 10^{-1}$,
 $a_{11}=1.10302752\cdot 10^{-4}$, $a_{12}=9.04905679\cdot 10^{-6}$, $a_{13}=1.04605847$, $a_{14}=8.06905161\cdot 10^{-1}$.

\begin{figure}
    \center{\includegraphics[width=0.97\linewidth]{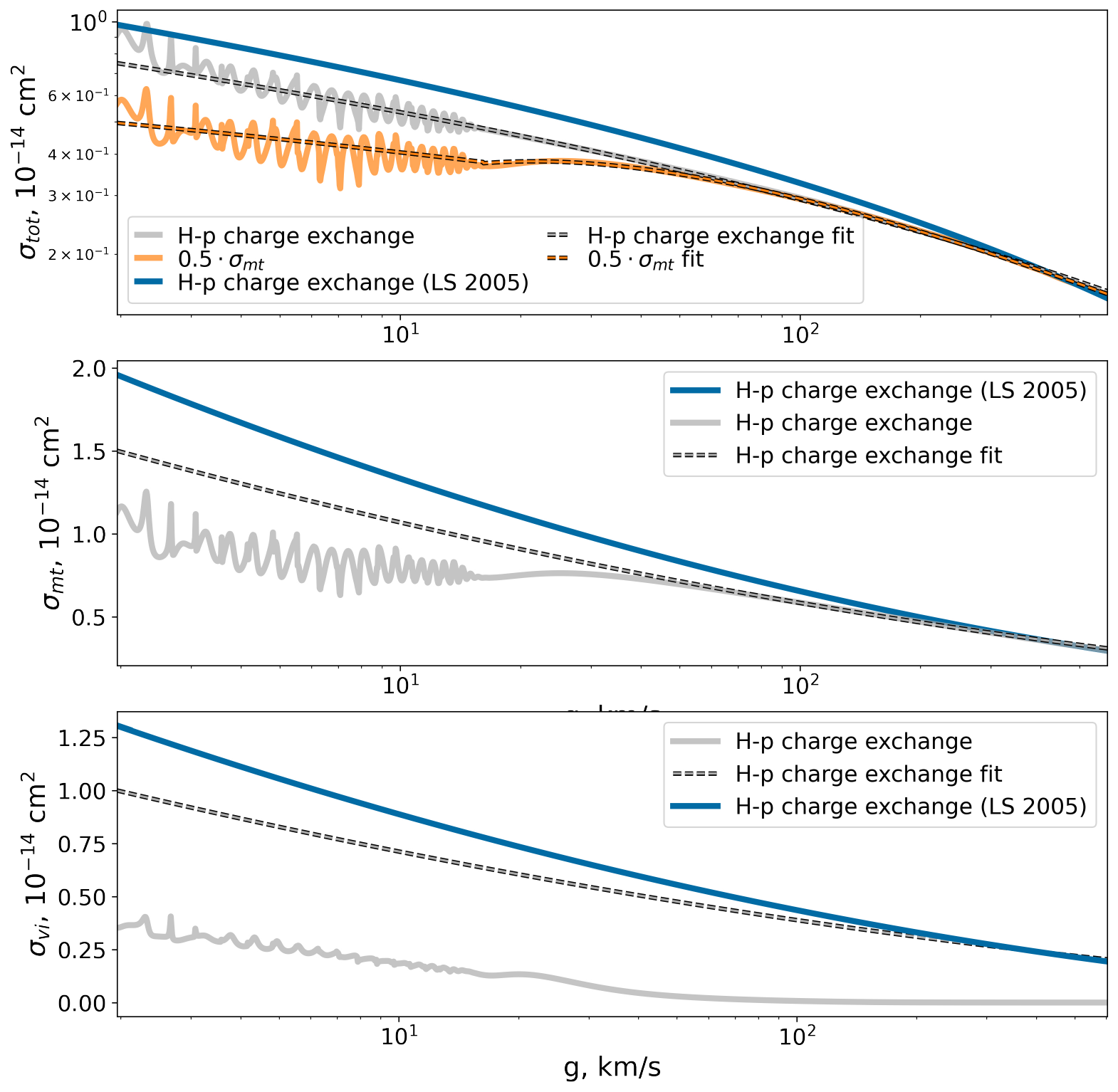}}
    \caption{ Comparison of charge exchange cross sections (total and momentum transfer): grey curve -- cross section by \cite{schultz_2016}, blue curve -- formula by \cite{lindsay_2005}, orange curve -- momentum transfer cross section divided by 2. Grey dashed curve is the cross section obtained in the $\chi=\pi$ assumption}
    \label{figure: charge_exchange_cs}
\end{figure}

\subsection{Charge-exchange cross section} \label{appendix: ch ex cross section}
\begin{figure}
    \center{\includegraphics[width=1\linewidth]{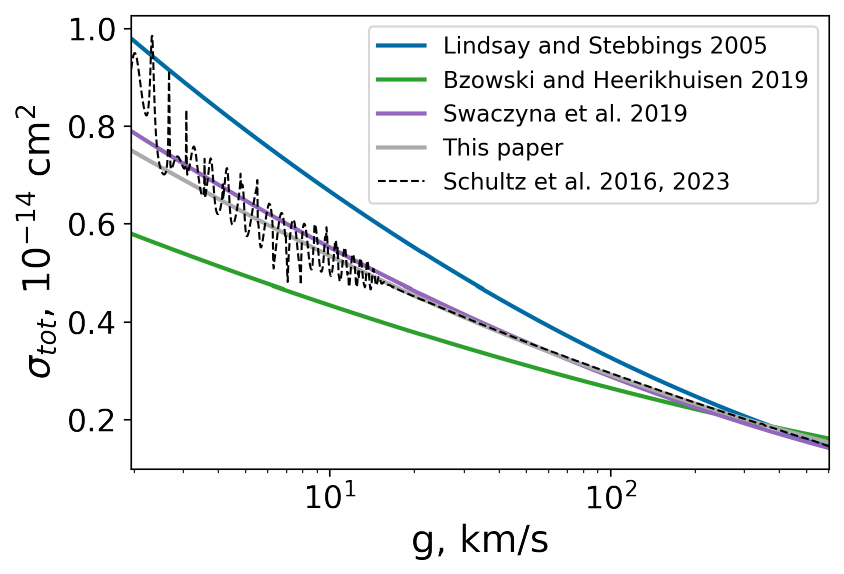}}
    \caption{ Comparison of total charge exchange cross section approximations: blue curve -- approximation by \cite{lindsay_2005}, green curve -- fit by \cite{Bzowski_2020}, violet curve -- fit by \cite{swaczyna2019}, grey curve -- fit used in this paper. Black dashed curve is the cross section from \cite{schultz_2016, schultz2023}}
    \label{figure: cs_comp2}
\end{figure}

In the numerous number of previous studies the charge exchange process has been considered under simplified assumption when the velocities of interacting particles are assumed to be unchangeable during the interaction, i.e. charge exchange was considered as a jump of electron from H atom to proton without any other changes. 
In the inertial rest frame connected with the center of mass of colliding particles this means the scattering at the angle of $\chi = \pi$. Therefore, this approach we call $\chi = \pi$ assumption, throughout this paper.
 With this assumption the differential charge exchange cross section can be written as 
\begin{equation}\label{sigma_ex_diff}
\frac{d\sigma_{ex}}{d \Omega} (g, \chi) = \frac{1}{4 \pi}\sigma_{tot,ex} \delta(\chi-\pi),
\end{equation}
where $\delta(x)$ is the delta-function.

The upper panel of Figure \ref{figure: diff_cs} shows
the differential charge exchange cross section at three different energies.
Although ultimate maxima of the cross sections are seen at $\chi = \pi$ for all energies, scattering at the angles less than $\pi$ appears as well. The less is the energy of collision the larger is the deflection from  $\chi = \pi$ may appear. (The scattering at the angles different from $\chi = \pi$ during charge exchange we call "charge exchange angle scattering" throughout the paper for shortness). 

To evaluate possible effects of the charge exchange angle scattering compare the momentum transfer and viscosity cross sections calculated from the differential cross sections by using equations (\ref{sigma_mt}) - (\ref{sigma_vi}) with
with those calculated under  $\chi = \pi$ assumption. In the latter case $\sigma_{mt, ex} = 2 \sigma_{tot,ex}$, $\sigma_{vi, ex} = \frac{4}{3} \sigma_{tot,ex}$, where the equation (\ref{sigma_tot_ex}) has been used for $\sigma_{tot,ex}$

Comparison in the middle panel of Figure \ref{figure: charge_exchange_cs} shows that for the relative velocities more than $\sim$70 km/s the two charge exchange cross section coincides. The difference becomes significant at smaller velocities. At the velocities less than 10 km/s the different increases up to 50\%.  

For $\sigma_{vi}$ the difference is prominent even at high velocities where the angle dependent cross section values drop almost to zero. At low velocities ($g<10$) the values differ by a factor of 4.

Besides, we present a comparison with the cross section by \cite{lindsay_2005} widely used in the modeling of the heliosphere. It is important to mention that this cross section was obtained for a range of velocities starting with $\sim 30$ km/s, thus, it may not show the correct results at lower velocities. \cite{schultz_2016} compared their theoretical calculations with the experiment data and \cite{lindsay_2005} results at high velocities (30-2000 km/s) and claimed that the overall agreement is quite good. As it can be seen from the Figure \ref{figure: charge_exchange_cs} for lower velocities Lindsay and Stebbings cross section is higher than the theoretical one calculated by \cite{schultz_2016}.
\begin{figure*}[t]
    \center{\includegraphics[width=1\linewidth]{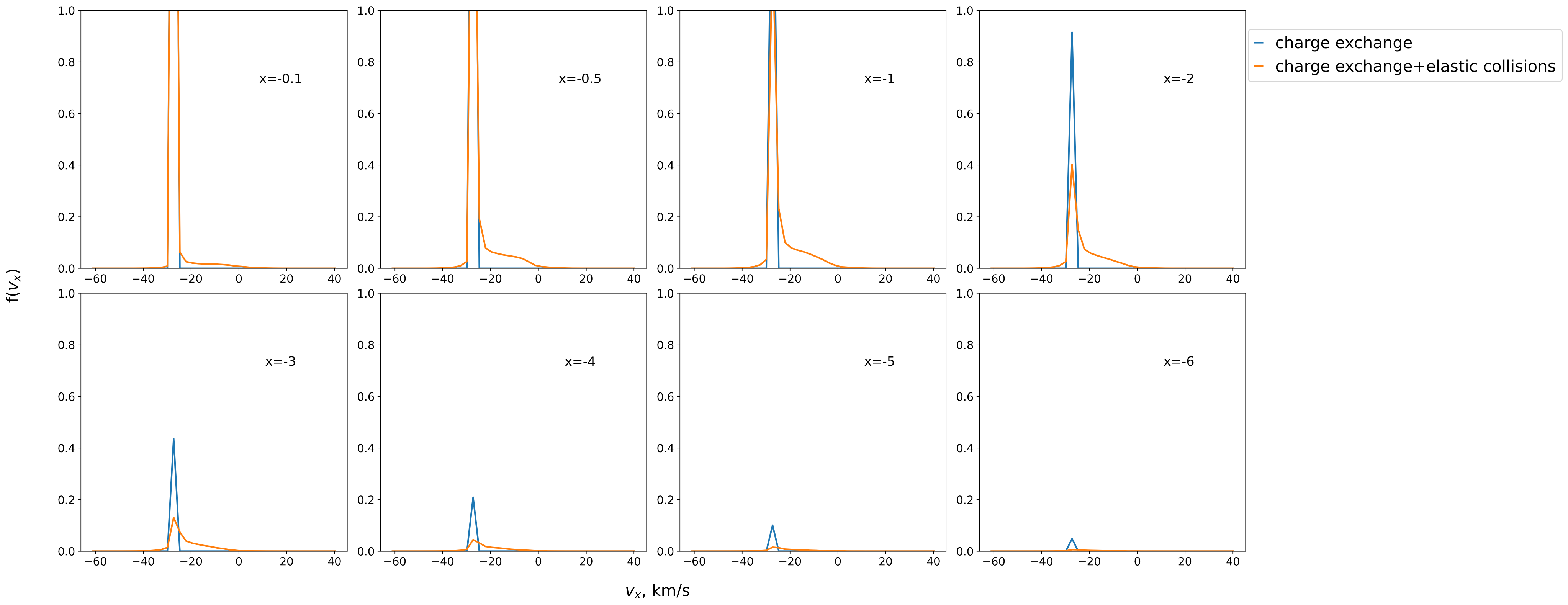}}
    \caption{
    The $v_{x}$-projection of the velocity distribution function for various normalized distances ($x=L/l_{0}$) in the homgenic plasma layer. The blue curves represent the evolution of the distribution function when only charge exchange with scattering angle $\chi=\pi$ is considered, while the orange curves include both charge exchange and H-p elastic collisions.}
      
    \label{figure: test}
\end{figure*}

Since the \cite{lindsay_2005} formula is only applicable within a limited range, efforts have been made to develop a more precise formula to model the charge exchange process. \cite{Bzowski_2020} in their work used a wider range of measurements from \cite{Barnett_1990}, spanning velocities from 4.79 to 368 km/s, and proposed a new fit. This fit is presented in Figure \ref{figure: cs_comp2} alongside the \cite{lindsay_2005} formula and the theoretical calculations by \cite{schultz_2016, schultz2023}. At low velocities, the \cite{Bzowski_2020} cross section is approximately 1.6 times lower than that of \cite{lindsay_2005}. The difference diminishes as the velocity increases, and the two curves intersect at velocities above 300 km/s. A similar pattern emerges when comparing the \cite{Bzowski_2020} cross section with the data from \cite{schultz_2016, schultz2023}. While the \cite{Bzowski_2020} cross section remains lower than the theoretical cross section at low velocities, the gap is less pronounced, as the theoretical cross section lies between the \cite{lindsay_2005} and \cite{Bzowski_2020} fits. At approximately 100 km/s, the \cite{Bzowski_2020} cross section aligns closely with the theoretical calculations.

\cite{swaczyna2019} for their calculations used the cross-section data from \cite{schultz_2016}, similar to the approach taken in this paper. However, for their fit, the authors employed a slightly modified formula:
\[
\sigma_\text{ex}(E_\text{proj}) = \left(4.049 - 0.447 \ln E \right)^2 
\times \left[1 - \exp\left(-\frac{60.5}{E}\right)\right]^{4.5} 
\times 10^{-16} \, \text{cm}^2.
\]
Here, $E$ is the projectile energy in KeV. As shown in Figure \ref{figure: cs_comp2}, this formula produces slightly higher cross section values at low velocities compared to the fit used in this paper. At higher velocities the results of both fits almost coincide.
\section{Test problem} \label{section:test_problem}

In order to understand the influence of H-p elastic collisions we performed additional calculations using a simplified toy model. In this model, hydrogen atoms travel through a flat layer of plasma with constant parameters: proton number density $n_{p}=1$ cm$^{-3}$, proton velocity $ V_{p}=-10$ km/s, and proton temperature $T_{p} =20000$ K.  The proton distribution function in the layer is maxwellian. The boundary conditions are set at $x=0$, where all hydrogen atoms have the same velocity $V_{H}=-26.4$ km/s, and the atom number density $n_{H}=1$ cm$^{-3}$. In this section, we focus on the evolution of primary hydrogen atoms, which are defined as atoms that have not undergone charge exchange. However, these primary atoms may still scatter through elastic collisions.

Figure \ref{figure: test} describes the evolution of the distribution function of primary H atoms with distance for two models. The distances presented in the figure are dimensionless: $x=L/l_{0}$, where $l_{0}$ is the mean free path for charge exchange process. The blue curve represents the model where only charge exchange with a scattering angle $\chi=\pi$ is considered. With the growth of the distance the peak of the distribution function decreases as more H atoms interact through charge exchange. At $x=-1$ the maximum is still high, but already at $x=-2$ it noticeably declines. At $x=-3$ the peak is almost twice smaller. Finally, at $x=-6$ only a small bump remains. The orange curve presents the model where H-p elastic collisions are taken into account in addition to charge exchange. The differences with the blue curve are noticed even at $x=-0.1$, where a broad zone of small velocity atoms appears. This zone is created due to the H-p elastic collisions. At $x=-0.5, -1$ this zone grows in size. Note that there are more atoms with velocities close to -26.4 km/s. This happens due to the form of the differential cross section of the H-p collisions: the differential cross section for H-p collisions peaks at small scattering angles, meaning that atoms are more likely to scatter at low angles, resulting in minimal changes to their velocity. At farther distances it can be noticed that the peak of the orange curve decreases faster than the blue one. At $x=-6$ the peak is almost indistinguishable. In summary, H-p elastic collisions accelerate the redistribution of hydrogen atom velocities and create a broad zone of atoms with intermediate velocities.

\end{document}